\documentclass[12pt,a4paper]{article}
\usepackage{booktabs}

\usepackage{ifthen} % for conditional statements
\newboolean{pdflatex}
\setboolean{pdflatex}{true} % False for eps figures 

\newboolean{articletitles}
\setboolean{articletitles}{true} % False removes titles in references

\newboolean{uprightparticles}
\setboolean{uprightparticles}{false} %True for upright particle symbols

\newboolean{inbibliography}
\setboolean{inbibliography}{false} %True once you enter the bibliography

\def\paperauthors{LHCb collaboration} % Leave as is for PAPER and CONF
\def\paperasciititle{\  Measurement of Xicc++ production \  
in pp collisions at 13 TeV} % Set ASCII title here
\def\papertitle{\  Measurement of $\Xiccpp$ production \  
in $pp$ collisions at $\sqrt{s}=13$~TeV} % Latex formatted title
\def\paperkeywords{{High Energy Physics}, {LHCb}} % Comma separated list
\def\papercopyright{\the\year\ CERN on behalf of the LHCb collaboration}
\def\paperlicence{CC-BY-4.0 licence}
\def\paperlicenceurl{https://creativecommons.org/licenses/by/4.0/}

\RequirePackage[normalem]{ulem} %DIF PREAMBLE
\RequirePackage{color}\definecolor{RED}{rgb}{1,0,0}\definecolor{BLUE}{rgb}{0,0,1} %DIF PREAMBLE
\usepackage[top=1in, bottom=1.25in, left=1in, right=1in]{geometry}

\usepackage{microtype}
\usepackage{lineno}  % for line numbering during review
\usepackage{xspace} % To avoid problems with missing or double spaces after
\usepackage{caption} %these three command get the figure and table captions automatically small

\usepackage{graphicx}  % to include figures (can also use other packages)
\usepackage{color}
\usepackage{colortbl}
\graphicspath{{./figs/}} % Make Latex search fig subdir for figures

\usepackage{amsmath} % Adds a large collection of math symbols
\usepackage{amssymb}
\usepackage{amsfonts}
\usepackage{upgreek} % Adds in support for greek letters in roman typeset

\newcommand*\patchAmsMathEnvironmentForLineno[1]{%
\expandafter\let\csname old#1\expandafter\endcsname\csname #1\endcsname
\expandafter\let\csname oldend#1\expandafter\endcsname\csname
end#1\endcsname
 \renewenvironment{#1}%
   {\linenomath\csname old#1\endcsname}%
   {\csname oldend#1\endcsname\endlinenomath}%
}
\newcommand*\patchBothAmsMathEnvironmentsForLineno[1]{%
  \patchAmsMathEnvironmentForLineno{#1}%
  \patchAmsMathEnvironmentForLineno{#1*}%
}
\AtBeginDocument{%
\patchBothAmsMathEnvironmentsForLineno{equation}%
\patchBothAmsMathEnvironmentsForLineno{align}%
\patchBothAmsMathEnvironmentsForLineno{flalign}%
\patchBothAmsMathEnvironmentsForLineno{alignat}%
\patchBothAmsMathEnvironmentsForLineno{gather}%
\patchBothAmsMathEnvironmentsForLineno{multline}%
\patchBothAmsMathEnvironmentsForLineno{eqnarray}%
}

\usepackage{hyperxmp}

\usepackage[pdftex,
            pdfauthor={\paperauthors},
            pdftitle={\paperasciititle},
            pdfkeywords={\paperkeywords},
            pdfcopyright={Copyright (C) \papercopyright},
            pdflicenseurl={\paperlicenceurl}]{hyperref}

\usepackage[all]{hypcap} % Internal hyperlinks to floats.

\usepackage{xspace} 
\usepackage{upgreek}

\def\lhcb {\mbox{LHCb}\xspace}

\def\babar  {\mbox{BaBar}\xspace}
\def\belle  {\mbox{Belle}\xspace}

\def\MagUp {\mbox{\em Mag\kern -0.05em Up}\xspace}

\ifthenelse{\boolean{uprightparticles}}%
{

 \def\Ppi         {\ensuremath{\uppi}\xspace}

 \def\Ppsi        {\ensuremath{\uppsi}\xspace}

 \def\PDelta      {\ensuremath{\Delta}\xspace}                 
 \def\PXi      {\ensuremath{\Xi}\xspace}                 
 \def\PLambda      {\ensuremath{\Lambda}\xspace}                 
 \def\PSigma      {\ensuremath{\Sigma}\xspace}                 
 \def\POmega      {\ensuremath{\Omega}\xspace}                 
 \def\PUpsilon      {\ensuremath{\Upsilon}\xspace}                 
 
 \def\PB      {\ensuremath{\mathrm{B}}\xspace}                 
                  
 \def\PD      {\ensuremath{\mathrm{D}}\xspace}

 \def\PJ      {\ensuremath{\mathrm{J}}\xspace}                 
 \def\PK      {\ensuremath{\mathrm{K}}\xspace}

 \def\Pb      {\ensuremath{\mathrm{b}}\xspace}                 
 \def\Pc      {\ensuremath{\mathrm{c}}\xspace}

 \def\Pi      {\ensuremath{\mathrm{i}}\xspace}

 \def\Pp      {\ensuremath{\mathrm{p}}\xspace}

}
{

 \def\Ppi         {\ensuremath{\pi}\xspace}

 \def\Ppsi        {\ensuremath{\psi}\xspace}                 
                  
 \mathchardef\PDelta="7101
 \mathchardef\PXi="7104
 \mathchardef\PLambda="7103
 \mathchardef\PSigma="7106
 \mathchardef\POmega="710A
 \mathchardef\PUpsilon="7107
                  
 \def\PB      {\ensuremath{B}\xspace}                 
                  
 \def\PD      {\ensuremath{D}\xspace}

 \def\PJ      {\ensuremath{J}\xspace}                 
 \def\PK      {\ensuremath{K}\xspace}

 \def\Pb      {\ensuremath{b}\xspace}                 
 \def\Pc      {\ensuremath{c}\xspace}

 \def\Pi      {\ensuremath{i}\xspace}

 \def\Pp      {\ensuremath{p}\xspace}

}

\makeatletter
\ifcase \@ptsize \relax% 10pt
  \newcommand{\miniscule}{\@setfontsize\miniscule{4}{5}}% \tiny: 5/6
\or% 11pt
  \newcommand{\miniscule}{\@setfontsize\miniscule{5}{6}}% \tiny: 6/7
\or% 12pt
  \newcommand{\miniscule}{\@setfontsize\miniscule{5}{6}}% \tiny: 6/7
\fi
\makeatother

\DeclareRobustCommand{\optbar}[1]{\shortstack{{\miniscule (\rule[.5ex]{1.25em}{.18mm})}
  \\ [-.7ex] $#1$}}

\def\cquark    {{\ensuremath{\Pc}}\xspace}

\def\bquark    {{\ensuremath{\Pb}}\xspace}

\def\pion   {{\ensuremath{\Ppi}}\xspace}

\def\pip    {{\ensuremath{\pion^+}}\xspace}
\def\pim    {{\ensuremath{\pion^-}}\xspace}

\def\kaon    {{\ensuremath{\PK}}\xspace}
  \def\Kbar    {{\kern 0.2em\overline{\kern -0.2em \PK}{}}\xspace}

\def\KorKbar    {\kern 0.18em\optbar{\kern -0.18em K}{}\xspace}

\def\Km      {{\ensuremath{\kaon^-}}\xspace}

  \def\Dbar    {{\kern 0.2em\overline{\kern -0.2em \PD}{}}\xspace}
\def\D       {{\ensuremath{\PD}}\xspace}

\def\DorDbar    {\kern 0.18em\optbar{\kern -0.18em D}{}\xspace}

\def\Dp      {{\ensuremath{\D^+}}\xspace}

\def\B       {{\ensuremath{\PB}}\xspace}
\def\Bbar    {{\ensuremath{\kern 0.18em\overline{\kern -0.18em \PB}{}}}\xspace}

\def\BorBbar    {\kern 0.18em\optbar{\kern -0.18em B}{}\xspace}

\def\Bc      {{\ensuremath{\B_\cquark^+}}\xspace}

\def\jpsi     {{\ensuremath{{\PJ\mskip -3mu/\mskip -2mu\Ppsi\mskip 2mu}}}\xspace}

  \def\Y#1S{\ensuremath{\PUpsilon{(#1S)}}\xspace}% no space before {...}!

\def\proton      {{\ensuremath{\Pp}}\xspace}

\def\Xires       {{\ensuremath{\PXi}}\xspace}

\def\Lz          {{\ensuremath{\PLambda}}\xspace}
\def\Lbar        {{\ensuremath{\kern 0.1em\overline{\kern -0.1em\PLambda}}}\xspace}
\def\LorLbar    {\kern 0.18em\optbar{\kern -0.18em \PLambda}{}\xspace}

\def\Lb      {{\ensuremath{\Lz^0_\bquark}}\xspace}

\def\Lc      {{\ensuremath{\Lz^+_\cquark}}\xspace}

\def\Xicp    {{\ensuremath{\Xires^+_\cquark}}\xspace}

\def\BF         {{\ensuremath{\mathcal{B}}}\xspace}

\def\BR         {\BF}
\newcommand{\decay}[2]{\ensuremath{#1\!\to #2}\xspace}         % {\Pa}{\Pb \Pc}

\def\to                 {\ensuremath{\rightarrow}\xspace}

\def\AT#1     {\ensuremath{A_{\mathrm{T}}^{#1}}\xspace}           % 2

\def\C#1      {\ensuremath{\mathcal{C}_{#1}}\xspace}                       % 9
\def\Cp#1     {\ensuremath{\mathcal{C}_{#1}^{'}}\xspace}                    % 7
\def\Ceff#1   {\ensuremath{\mathcal{C}_{#1}^{\mathrm{(eff)}}}\xspace}        % 9  
\def\Cpeff#1  {\ensuremath{\mathcal{C}_{#1}^{'\mathrm{(eff)}}}\xspace}       % 7
\def\Ope#1    {\ensuremath{\mathcal{O}_{#1}}\xspace}                       % 2
\def\Opep#1   {\ensuremath{\mathcal{O}_{#1}^{'}}\xspace}                    % 7

\newcommand{\tev}{\ifthenelse{\boolean{inbibliography}}{\ensuremath{~T\kern -0.05em eV}}{\ensuremath{\mathrm{\,Te\kern -0.1em V}}}\xspace}
\newcommand{\gev}{\ensuremath{\mathrm{\,Ge\kern -0.1em V}}\xspace}
\newcommand{\mev}{\ensuremath{\mathrm{\,Me\kern -0.1em V}}\xspace}
\newcommand{\kev}{\ensuremath{\mathrm{\,ke\kern -0.1em V}}\xspace}
\newcommand{\ev}{\ensuremath{\mathrm{\,e\kern -0.1em V}}\xspace}
\newcommand{\gevc}{\ensuremath{{\mathrm{\,Ge\kern -0.1em V\!/}c}}\xspace}
\newcommand{\mevc}{\ensuremath{{\mathrm{\,Me\kern -0.1em V\!/}c}}\xspace}
\newcommand{\gevcc}{\ensuremath{{\mathrm{\,Ge\kern -0.1em V\!/}c^2}}\xspace}
\newcommand{\gevgevcccc}{\ensuremath{{\mathrm{\,Ge\kern -0.1em V^2\!/}c^4}}\xspace}
\newcommand{\mevcc}{\ensuremath{{\mathrm{\,Me\kern -0.1em V\!/}c^2}}\xspace}

\def\mum  {\ensuremath{{\,\upmu\mathrm{m}}}\xspace}

\def\nb {\ensuremath{\mathrm{ \,nb}}\xspace}

\def\invfb   {\ensuremath{\mbox{\,fb}^{-1}}\xspace}

\def\ps   {\ensuremath{{\mathrm{ \,ps}}}\xspace}

\newcommand{\stat}{\ensuremath{\mathrm{\,(stat)}}\xspace}
\newcommand{\syst}{\ensuremath{\mathrm{\,(syst)}}\xspace}

\newcommand{\chisq}{\ensuremath{\chi^2}\xspace}

\newcommand{\chisqip}{\ensuremath{\chi^2_{\text{IP}}}\xspace}

\def\gsim{{~\raise.15em\hbox{$>$}\kern-.85em
          \lower.35em\hbox{$\sim$}~}\xspace}
\def\lsim{{~\raise.15em\hbox{$<$}\kern-.85em
          \lower.35em\hbox{$\sim$}~}\xspace}

\def\ptot       {\mbox{$p$}\xspace}
\def\pt         {\mbox{$p_{\mathrm{ T}}$}\xspace}

\def\mrad{\ensuremath{\mathrm{ \,mrad}}\xspace}

\def\evtgen     {\mbox{\textsc{EvtGen}}\xspace}

\def\geant      {\mbox{\textsc{Geant4}}\xspace}

\def\photos     {\mbox{\textsc{Photos}}\xspace}

\def\pythia     {\mbox{\textsc{Pythia}}\xspace}

\def\tell1  {TELL1\xspace}
\def\ukl1   {UKL1\xspace}

\newcommand{\eg}{\mbox{\itshape e.g.}\xspace}

\usepackage{cite} % Allows for ranges in citations
\usepackage{mciteplus}

\usepackage{xspace} 
\usepackage{upgreek}

\def\XiccCom{\ensuremath{2.22\, \pm 0.27\,\pm 0.29\,}}
\def\XiccTOS{\ensuremath{2.57\, \pm 0.51\,\pm 0.43\,}}
\def\XiccTIS{\ensuremath{2.11\, \pm 0.31\,\pm
    0.30\,}}
\def\totalLumi{\ensuremath{1.7\invfb}}

\def\XiccComl{\ensuremath{2.53\, \pm 0.30\,\pm 0.33\,}}
\def\XiccTOSl{\ensuremath{2.90\, \pm 0.57\,\pm 0.49\,}}
\def\XiccTISl{\ensuremath{2.41\, \pm 0.35\,\pm
    0.34\,}}
\def\XiccComh{\ensuremath{1.98\, \pm 0.23\,\pm 0.26\,}}
\def\XiccTOSh{\ensuremath{2.31\, \pm 0.46\,\pm 0.39\,}}
\def\XiccTISh{\ensuremath{1.88\, \pm 0.27\,\pm
    0.27\,}}

\def\DeltaOrDeltabar  {\kern 0.18em\optbar{\kern -0.18em \PDelta}{}\xspace}
\def\XiOrXibar        {\kern 0.18em\optbar{\kern -0.18em \PXi}{}\xspace}
\def\SigmaOrSigmabar  {\kern 0.18em\optbar{\kern -0.18em \PSigma}{}\xspace}
\def\OmegaOrOmegabar  {\kern 0.18em\optbar{\kern -0.18em \POmega}{}\xspace}

\def\Lcp          {{\ensuremath{\Lz^+_\cquark}}\xspace}

\def\Xicp         {{\ensuremath{\Xires^+_\cquark}}\xspace}

\def\Xiccpp       {{\ensuremath{\Xires^{++}_{\cquark\cquark}}}\xspace}

\def\Xiccp        {{\ensuremath{\Xires^{+}_{\cquark\cquark}}}\xspace}

\newcommand{\TeVnosp}{\ifthenelse{\boolean{inbibliography}}{\ensuremath{~T\kern -0.05em eV}}{\ensuremath{\mathrm{Te\kern -0.1em V}}}}
\newcommand{\GeVnosp}{\ensuremath{\mathrm{Ge\kern -0.1em V}}}
\newcommand{\MeVnosp}{\ensuremath{\mathrm{Me\kern -0.1em V}}}
\newcommand{\keVnosp}{\ensuremath{\mathrm{ke\kern -0.1em V}}}
\newcommand{\eVnosp}{\ensuremath{\mathrm{e\kern -0.1em V}}}
\newcommand{\GeVcnosp}{\ensuremath{{\mathrm{Ge\kern -0.1em V\!/}c}}}
\newcommand{\MeVcnosp}{\ensuremath{{\mathrm{Me\kern -0.1em V\!/}c}}}
\newcommand{\GeVccnosp}{\ensuremath{{\mathrm{Ge\kern -0.1em V\!/}c^2}}}
\newcommand{\MeVccnosp}{\ensuremath{{\mathrm{Me\kern -0.1em V\!/}c^2}}}
\newcommand{\GeVGeVccccnosp}{\ensuremath{{\mathrm{Ge\kern -0.1em V^2\!/}c^4}}}

\def\genxicctwo   {\mbox{\textsc{GenXicc2.0}}\xspace}

\usepackage{longtable} % only for template; not usually to be used in PAPERs
\usepackage{multirow}

\def\to                 {\ensuremath{\rightarrow}\xspace}
\def\Lc      {{\ensuremath{\Lz^+_\cquark}}\xspace}
\def\Xiccpp       {{\ensuremath{\Xires^{++}_{\cquark\cquark}}}\xspace}

\begin{document}

%%%%%%%%%%%%%%%%%%%%%%%%%
%%%%% Title     %%%%%%%%%
%%%%%%%%%%%%%%%%%%%%%%%%%
\renewcommand{\thefootnote}{\fnsymbol{footnote}}
\setcounter{footnote}{1}

% %%%%%%% CHOOSE TITLE PAGE--------
%\onecolumn
% $Id: title-LHCb-PAPER.tex 107952 2017-05-17 11:30:35Z uegede $
% ===============================================================================
% Purpose: LHCb-PAPER journal paper title page template
% Author: 
% Created on: 2010-09-25
% ===============================================================================

%%%%%%%%%%%%%%%%%%%%%%%%%
%%%%%  TITLE PAGE  %%%%%%
%%%%%%%%%%%%%%%%%%%%%%%%%
\begin{titlepage}
\pagenumbering{roman}

% Header ---------------------------------------------------
\vspace*{-1.5cm}
\centerline{\large EUROPEAN ORGANIZATION FOR NUCLEAR RESEARCH (CERN)}
\vspace*{1.5cm}
\noindent
\begin{tabular*}{\linewidth}{lc@{\extracolsep{\fill}}r@{\extracolsep{0pt}}}
\ifthenelse{\boolean{pdflatex}}% Logo format choice
{\vspace*{-2.7cm}\mbox{\!\!\!\includegraphics[width=.14\textwidth]{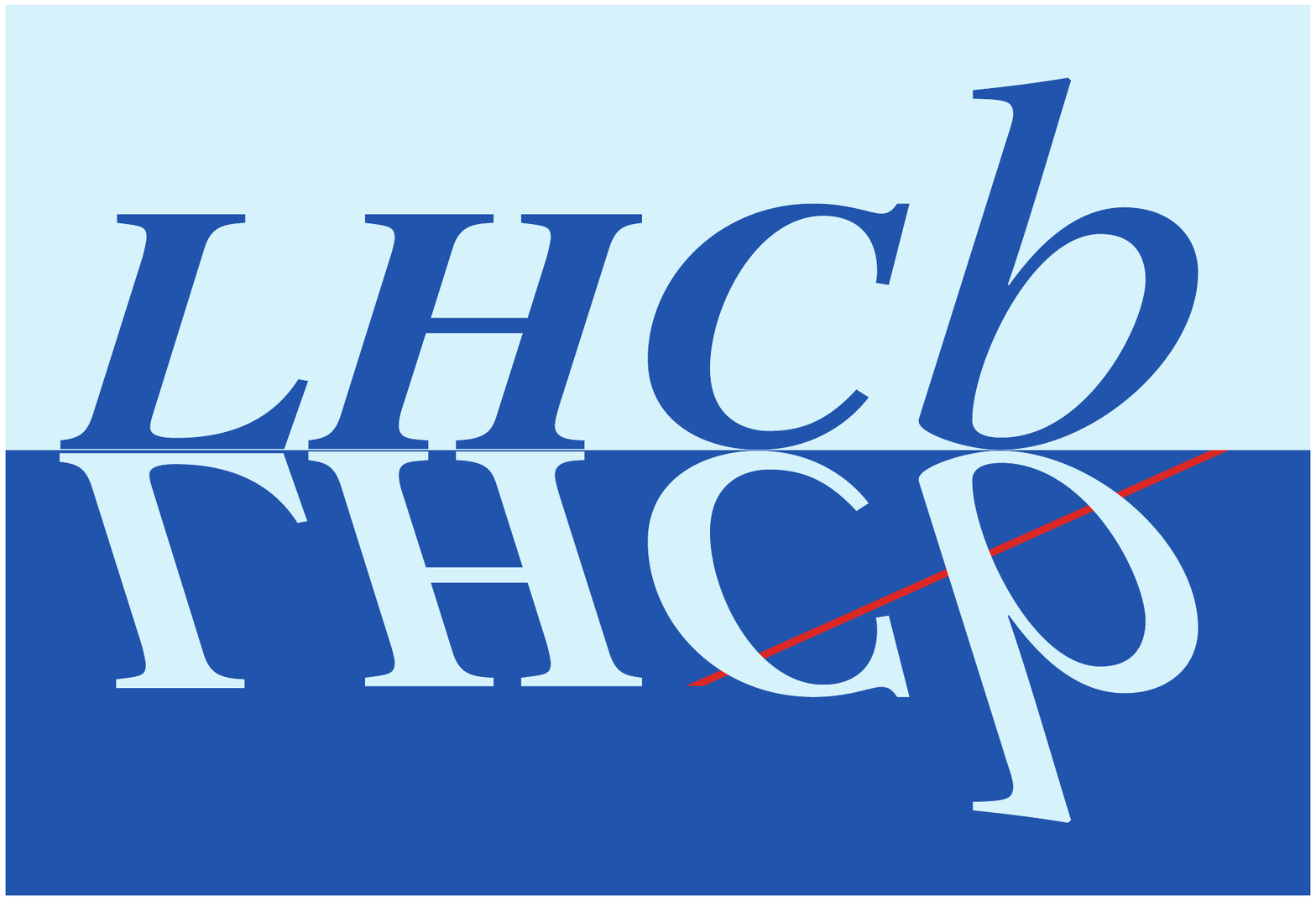}} & &}%
{\vspace*{-1.2cm}\mbox{\!\!\!\includegraphics[width=.12\textwidth]{lhcb-logo.eps}} & &}%
\\
 & & CERN-EP-2019-220 \\  % ID 
 & & LHCb-PAPER-2019-035 \\  % ID 
 & & 4 February 2020 \\ % Date - Can also hardwire e.g.: 23 March 2010
% not in paper \hline
\end{tabular*}

\vspace*{4.0cm}

% Title --------------------------------------------------
{\normalfont\bfseries\boldmath\huge
\begin{center}
% DO NOT EDIT HERE. Instead edit macro in main.tex to keep metadata correct
  \papertitle 
\end{center}
}

\vspace*{2.0cm}

% Authors -------------------------------------------------
\begin{center}
%In the footnote, replace 'paper' by 'Letter' in case of submission to PRL or PLB 
% Edit macro in main.tex to keep metadata correct
\paperauthors\footnote{Authors are listed at the end of this paper.}
\end{center}

\vspace{\fill}

% Abstract -----------------------------------------------
\begin{abstract}
  \noindent
  The production of $\Xiccpp$ baryons in proton-proton collisions at a centre-of-mass energy of $\sqrt{s}=13$~TeV is measured in the transverse-momentum range $4<\pt<15\gevc$ and the rapidity range $2.0<y<4.5$. The data used in this measurement correspond to an integrated luminosity of 1.7\invfb, recorded by the LHCb experiment during 2016.
  The ratio of the $\Xiccpp$ production cross-section times 
  the branching fraction of the 
  $\Xiccpp \to \Lc K^-\pi^+ \pi^+$ decay relative to 
  the prompt $\Lc$ production
  cross-section is found to be $(2.22\pm 0.27 \pm 0.29)\times 10^{-4}$, assuming the central value of the measured $\Xiccpp$ lifetime, where the first uncertainty
  is statistical and the second systematic.

\end{abstract}

\vspace*{2.0cm}

\begin{center}
  Published in Chin. Phys. C44 (2020) 022001
\end{center}

\vspace{\fill}

{\footnotesize 
% Edit macro in main.tex to keep metadata correct
\centerline{\copyright~\papercopyright, license \href{\paperlicenceurl}{\paperlicence}.}}
\vspace*{2mm}

\end{titlepage}

%%%%%%%%%%%%%%%%%%%%%%%%%%%%%%%%
%%%%%  EOD OF TITLE PAGE  %%%%%%
%%%%%%%%%%%%%%%%%%%%%%%%%%%%%%%%

%  empty page follows the title page ----
\newpage
\setcounter{page}{2}
\mbox{~}
%\newpage

\cleardoublepage

%\twocolumn
% %%%%%%%%%%%%% ---------

\renewcommand{\thefootnote}{\arabic{footnote}}
\setcounter{footnote}{0}

\pagestyle{plain} % restore page numbers for the main text
\setcounter{page}{1}
\pagenumbering{arabic}

\section{Introduction}
\label{sec:Introduction}
The quark model~\cite{GellMann:1964nj,Zweig:352337} predicts the existence of multiplets
of baryon and meson states.
Baryons containing two charm quarks and a light quark provide a unique
system for testing the low-energy limit of   
quantum chromodynamics (QCD).
The production of doubly charmed baryons 
at hadron colliders 
can be 
treated as two independent processes:
production of a $cc$ diquark followed by the hadronisation of the diquark into a baryon~\cite{Berezhnoy:1998aa,kiselev2002baryons,Ma:2003zk,Chang:2006xp,Chang:2006eu,Zhang:2011hi,Chang:2005bf}. 
The production cross-section of doubly charmed baryons in
proton-proton collisions at a centre-of-mass energy $\sqrt{s}=13\tev$ is predicted to be 
in the range
60--1800\nb~\cite{Berezhnoy:1998aa,kiselev2002baryons,Ma:2003zk,Chang:2006xp,Chang:2006eu,Zhang:2011hi,Chang:2005bf},
which is between 
$10^{-4}$ and $10^{-3}$ times that of the total charm production~\cite{kiselev2002baryons}. 

A doubly charmed baryon was first reported by the SELEX collaboration~\cite{Mattson:2002vu,Ocherashvili:2004hi}. They found that 20$\%$ of their $\Lcp$ yield originated from \Xiccp decays, which is several orders of magnitude higher than theoretical prediction~\cite{kiselev2002baryons}. However, this signal has not been confirmed by searches performed at the FOCUS~\cite{Ratti:2003ez}, \babar~\cite{Aubert:2006qw}, \belle~\cite{Chistov:2006zj}, and \lhcb~\cite{LHCb-PAPER-2013-049, LHCb-PAPER-2019-029} experiments.
Recently, the LHCb collaboration
observed a peak in the $\Lcp\Km\pip\pip$ mass spectrum at a mass of
$3621.40 \pm 0.78\mevcc$~\cite{LHCb-PAPER-2017-018}, 
consistent with expectations for the $\Xiccpp$ baryon.
The $\Xiccpp$ lifetime was measured to be 
$0.256^{+0.024}_{-0.022} \stat \pm 0.014\syst\ps$~\cite{LHCb-PAPER-2018-019}, indicating that it decays through the weak interaction.
A new decay mode, $\Xiccpp \to \Xicp \pip$, was observed by  the LHCb collaboration~\cite{LHCb-PAPER-2017-040}, and the measured $\Xiccpp$ mass was found to be
consistent with that measured using $\Xiccpp \to
\Lcp\Km\pip\pip$ decays.
The $\Xiccpp\to\Dp\proton\Km\pip$ decay has been searched for,
but no signal was found~\cite{LHCb-PAPER-2019-011}.

This paper presents a measurement of  
\Xiccpp production  
in $pp$ collisions at a centre-of-mass energy of $\sqrt{s}=13\tev$, 
following the same analysis strategy as that used in Refs.~\cite{LHCb-PAPER-2013-049, LHCb-PAPER-2017-018,LHCb-PAPER-2018-019}.
The $\Xiccpp$ production cross-section, 
$\sigma(\Xiccpp)$, times the branching fraction of the 
\mbox{\decay{\Xiccpp}{\Lcp\Km\pip\pip}} decay, is measured relative to the prompt $\Lc$ production cross-section, $\sigma(\Lc)$, 
in the transverse momentum range 
$4<\pt<15\gevc$ and the rapidity range $2.0<y<4.5$.
The data used correspond to an integrated luminosity of $\totalLumi$
collected by the LHCb experiment in 2016. The $\Lc$ baryon is reconstructed via
the $\Lc \to pK^-\pip$ decay. 
The inclusion of the charge-conjugate decay processes is implied throughout this paper.
The production rate ratio is defined as,
\begin{equation}
R \equiv \frac{\sigma(\Xiccpp)\times\BR(\decay{\Xiccpp}{\Lcp\Km\pip\pip}) }{\sigma(\Lc) } = 
\frac{ 
  N_{\text{sig}}
}{
  N_{\text{norm}}
}
\frac{ 
  \varepsilon_{\text{norm}}
}{
  \varepsilon_{\text{sig}}
},
\label{eq:strategy:defineMainResult}
\end{equation}
where
``sig'' and ``norm'' refer to the signal (\Xiccpp) and normalisation (\Lc) modes,
$N$ is the signal yield and $\varepsilon$ is the total efficiency to reconstruct and select these decays.

\section{Detector and simulation}
\label{sec:Detector}
The \lhcb detector~\cite{LHCb-DP-2008-001,LHCb-DP-2014-002} is a single-arm forward
spectrometer covering the \mbox{pseudorapidity} range $2<\eta <5$,
designed for the study of particles containing \bquark or \cquark
quarks. The detector includes a high-precision tracking system
consisting of a silicon-strip vertex detector surrounding the $pp$
interaction region~\cite{LHCb-DP-2014-001}, a large-area silicon-strip detector located
upstream of a dipole magnet with a bending power of about
$4{\mathrm{\,Tm}}$, and three stations of silicon-strip detectors and straw
drift tubes~\cite{LHCb-DP-2017-001} placed downstream of the magnet.
The tracking system provides a measurement of the momentum, \ptot, of charged particles with
a relative uncertainty that varies from 0.5\% at low momentum to 1.0\% at 200\gevc.
The minimum distance of a track to a primary vertex, the impact parameter,
is measured with a resolution of $(15+29/\pt)\mum$, where $\pt$ is expressed in $\gevc$.
Different types of charged hadrons are distinguished using information
from two ring-imaging Cherenkov detectors~\cite{LHCb-DP-2012-003}.
Photons, electrons and hadrons are identified by a calorimeter system consisting of
scintillating-pad (SPD) and preshower detectors, an electromagnetic and a hadronic calorimeter. Muons are identified by a
system composed of alternating layers of iron and multiwire
proportional chambers~\cite{LHCb-DP-2012-002}.
The online event selection is performed by a trigger~\cite{LHCb-DP-2012-004}, 
which consists of a hardware stage, based on information from the calorimeters and muon
systems~\cite{LHCb-DP-2013-004, LHCb-DP-2013-001}, followed by a software stage, which applies a full event
reconstruction incorporating near-real-time
alignment and calibration of the
detector~\cite{LHCb-DP-2019-001}.
The output of the reconstruction performed in the software trigger~\cite{LHCb-DP-2016-001}
is used as input to the present analysis.

Simulated samples are required to develop the candidate selection and
to estimate the efficiency of the detector acceptance and the
imposed selection requirements.
Simulated $pp$ collisions are generated using
\pythia~\cite{Sjostrand:2007gs,*Sjostrand:2006za} 
with a specific \lhcb configuration~\cite{LHCb-PROC-2010-056}.  
A dedicated package, \genxicctwo~\cite{Chang:2009va},
is used to simulate the \Xiccpp baryon production.
Decays of unstable particles
are described by \evtgen~\cite{Lange:2001uf}, in which final-state
radiation is generated using \photos~\cite{Golonka:2005pn}. The
interaction of the generated particles with the detector, and its response,
are simulated using the \geant
toolkit~\cite{Allison:2006ve, *Agostinelli:2002hh} as described in
Ref.~\cite{LHCb-PROC-2011-006}.

\section{Event selection}
\label{sec:Sel}
The  
$\Lcp \to \proton \Km \pip$ candidate is reconstructed through three charged particles identified as
\proton, \Km and \pip hadrons,
which form a common vertex and do not originate from any primary vertex (PV) in the event.
The decay vertex of the \Lcp candidate is required to be displaced from any PV
by requiring its proper decay time to be greater than
$0.15\ps$, corresponding to about 1.5 times the $\Lc$ decay time resolution~\cite{PDG2018}.
Each \Lcp candidate with mass in the range 2270--2306\mevcc
is then combined with three additional particles
  to form a \Xiccpp candidate.
The three particles must form a common vertex with the \Lcp candidate
  and have hadron-identification information
  consistent with them being two \pip mesons and one \Km meson.
The \Lc decay vertex is required to be downstream of the \Xiccpp vertex.
Additionally, the $\Xiccpp$ candidates must have $\pt > 4$\gevc
and originate from a PV.

The combinatorial background is suppressed using two multivariate classifiers
based on a boosted decision tree algorithm~\cite{Hocker:2007ht,*TMVA4}.
One classifier is optimised to select $\Lc$ candidates  irrespective of their origin, and the other is optimised to select $\Xiccpp$ candidates.
While both classifiers are applied to the signal channel, only the first is applied to the
normalisation decay channel.
The first classifier is trained with $\Lcp$ signal in the simulated $\Xiccpp$ sample
and background candidates in the $\Lcp$ mass sideband.
The second classifier is trained using data candidates in the $\Lcp$ and $\Xiccpp$ signal mass region, where wrong-sign (WS) $\Lcp K^-\pip\pim$ combinations are used as proxy for the background.
The first multivariate classifier is trained with the following
variables: 
the $\chi^2$  of the \Lc vertex fit;
the largest distance of closest approach 
among the decay products;
the scalar sum of the \pt and 
the smallest \pt of the three decay products of the \Lcp candidate;
the smallest and  largest $\chisqip$ of the decay products of the \Lcp candidate with respect to its PV.
Here, \chisqip is defined as the difference in $\chi^2$ of the PV fit
with and without the particle in question. The PV of any single particle is defined to be that with respect to which
the particle has the smallest \chisqip.
The second multivariate classifier is trained with the following
variables:
the $\chisqip$ of the \Xiccpp candidate to its PV;
the angle between the \Xiccpp momentum and the direction from the PV to the \Xiccpp decay vertex;
the logarithm of the $\chi^2$ of the $\Xiccpp$ flight distance
between the $\Xiccpp$ decay vertex and the PV;
the vertex fit $\chisq$ of the \Xiccpp candidate;
the $\chisq$ of a kinematic
refit~\cite{Hulsbergen:2005pu} that requires the \Xiccpp candidate to originate from a PV;
the scalar sum of the \pt and the
smallest \pt of the six final state tracks of the \Xiccpp candidate.
Here the flight distance $\chi^2$ is defined as the change in $\chi^2$
of the $\Xiccpp$ decay vertex if it is constrained to coincide with
the PV.
Candidates retained for analysis must have two classifier responses exceeding thresholds 
chosen by performing a two-dimensional maximisation of the figure of merit 
$\varepsilon/(5/2+\sqrt{B})$~\cite{Punzi:2003bu}.
Here $\varepsilon$ and $B$ are the estimated signal efficiency determined from signal simulation and
background yield under the signal peak, respectively. The background is estimated from the WS sample.
The same threshold of the first classifier, optimised for the signal mode, is applied to the normalisation mode.

Finally, the \Xiccpp and \Lcp candidates are required to have 
their transverse momentum and rapidity
in the fiducial ranges of 4--15\gevc and 2.0--4.5, respectively. 
After the multivariate selection is applied, 
events may still contain more than one $\Xiccpp$ candidate in the signal region. 
Candidates made of duplicate tracks are removed by
requiring all pairs of tracks with the same charge to have an opening angle larger than 
$0.5\mrad$.
Duplicate candidates,
which are due to the interchange between identical particles from the \Lcp 
decay or directly from the \Xiccpp decay 
(\eg, the \Km particle from the \Xiccpp decay and the \Km particle
from the \Lcp decay),
can cause peaking structures in the $\Xiccpp$ invariant mass distribution. In this case, one of the candidates
is chosen at random to be retained and the others are discarded.
The systematic uncertainty associated with this procedure is negligible.

\section{Signal yields}
\label{sec:signals}
After the full selection is applied, 
the data sets are further filtered into two disjoint subsamples using information 
from the hardware trigger.
The first contains candidates that are triggered by at least one of the 
$\Lc$ decay products with high transverse
energy deposited in the calorimeters, referred to as 
Triggered On Signal (TOS).
The second consists of the events that are 
exclusively triggered by particles unrelated to the signal decay products;
these events can, for example, be triggered by the decay products of the charmed hadrons
produced together with the signal baryon, 
referred to as 
exclusively Triggered Independently of Signal (exTIS).

\begin{figure}[!tbp]
\centering
\begin{minipage}[t]{0.49\textwidth}
\centering
\includegraphics[width=1.0\textwidth]{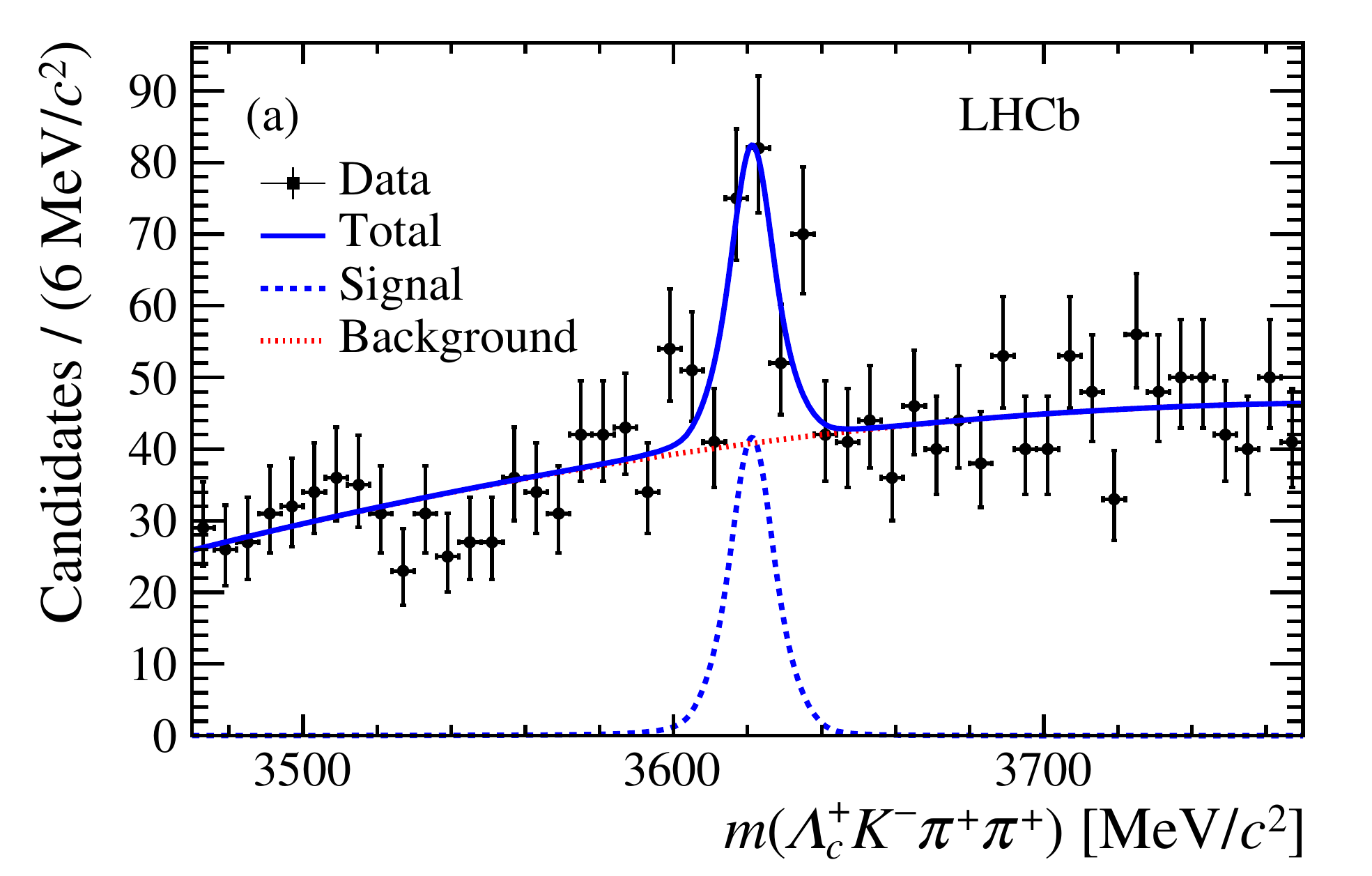}
\end{minipage}
\begin{minipage}[t]{0.49\textwidth}
\centering
\includegraphics[width=1.0\textwidth]{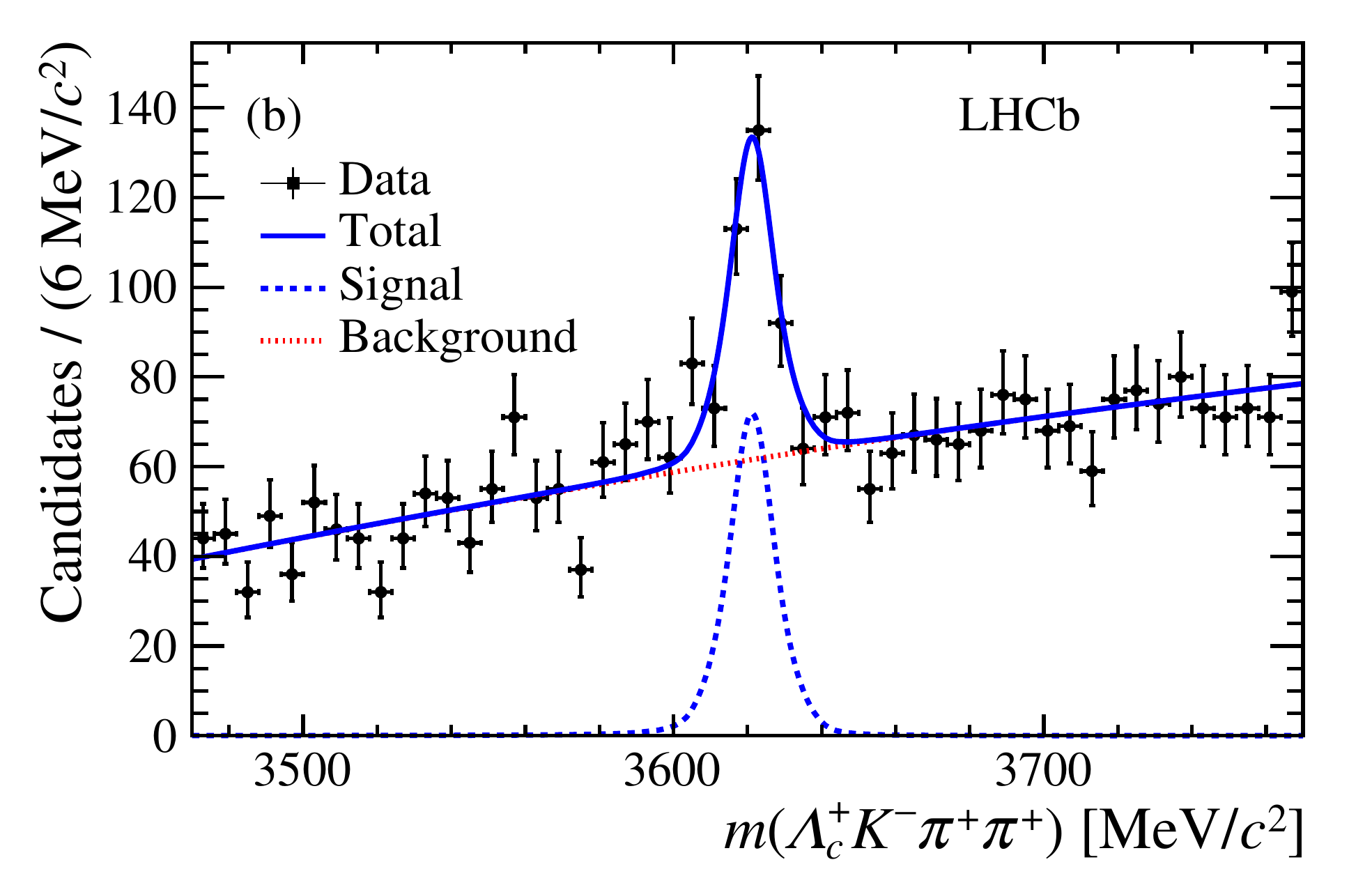}
\end{minipage}
\caption{Invariant-mass distributions of $\Xiccpp$ candidates (a) triggered by TOS and (b) triggered by exTIS,
with fit results shown.}
\label{fig:XiccMass}
\end{figure}

To determine the $\Xiccpp$ baryon signal yields, 
an unbinned extended maximum-likelihood
fit is performed simultaneously to
the $\Lcp K^- \pip\pip$ invariant-mass spectra 
in the interval 3470--3770\mevcc
of the two trigger
categories.
The mass distribution of the signal is described 
by the sum of 
a Gaussian function and a modified Gaussian function with power-law
tails on both sides of the function~\cite{Skwarnicki:1986xj} with a common peak position.
The tail parameters and the relative fraction of the two Gaussian functions for the signal model are determined from simulation,
while the
common peak position and the mass resolution
are allowed to vary in the fit.
The background is described by a second-order Chebyshev polynomial.
Figure~\ref{fig:XiccMass} shows 
the $\Lcp K^- \pip\pip$ invariant-mass distribution in data together with the 
fit results for the two
trigger categories.
The fit returns a mass of $3621.34 \pm 0.74 \mevcc$, 
and a mass resolution of $7.1 \pm 1.3 \mevcc$, where the uncertainties are statistical only.  

\begin{figure}[!tbp]
\centering
\begin{minipage}[t]{0.49\textwidth}
\centering
\includegraphics[width=1.0\textwidth]{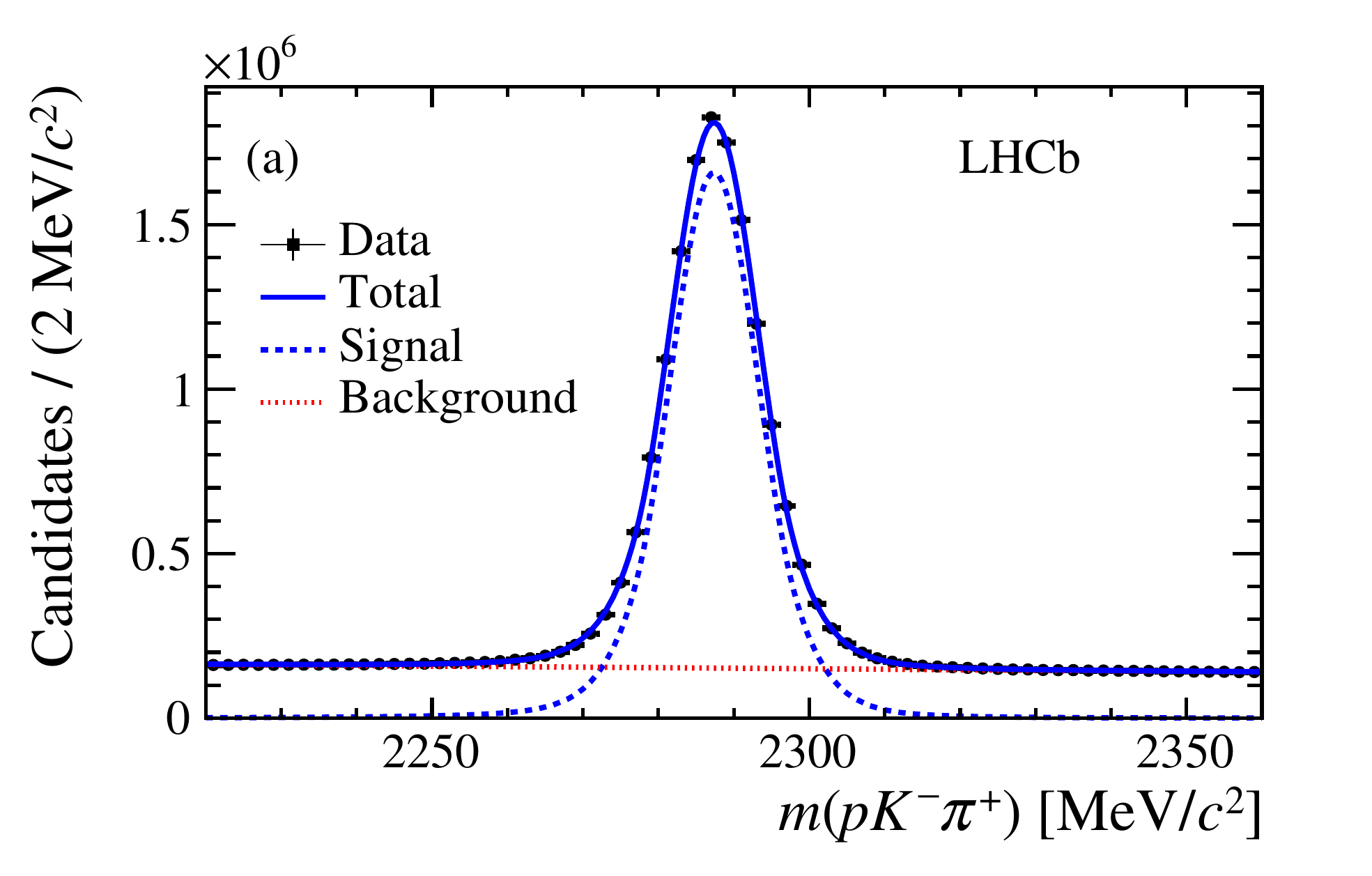}
\end{minipage}
\begin{minipage}[t]{0.49\textwidth}
\centering
\includegraphics[width=1.0\textwidth]{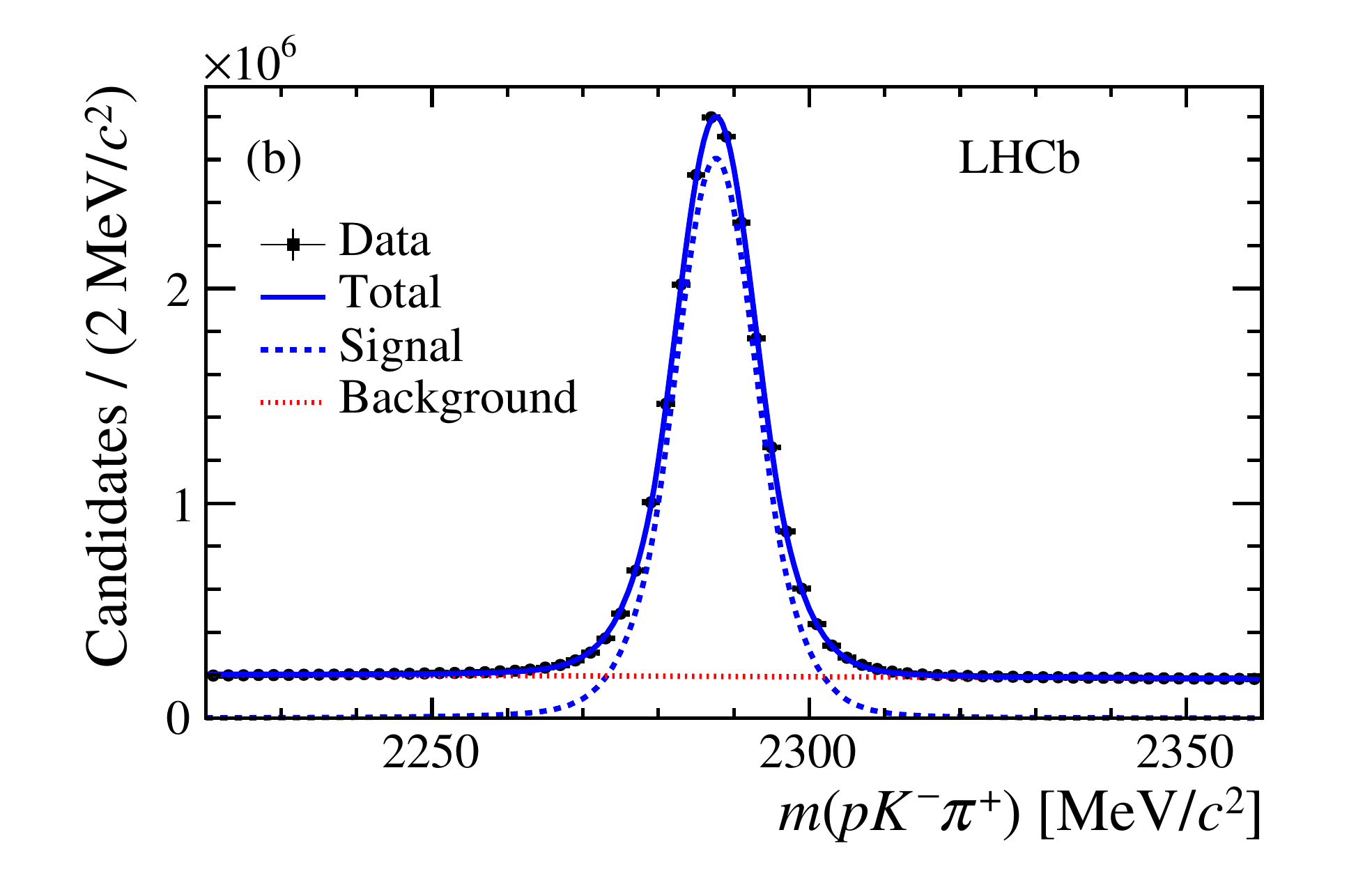}
\end{minipage}
\caption{Invariant-mass distributions of $\Lcp$ candidates (a)
triggered by TOS and (b) triggered by exTIS,
with fit results shown.}
\label{fig:LcMass}
\end{figure}

\begin{figure}[!tbp]
\centering
\begin{minipage}[t]{0.49\textwidth}
\centering
\includegraphics[width=1.0\textwidth]{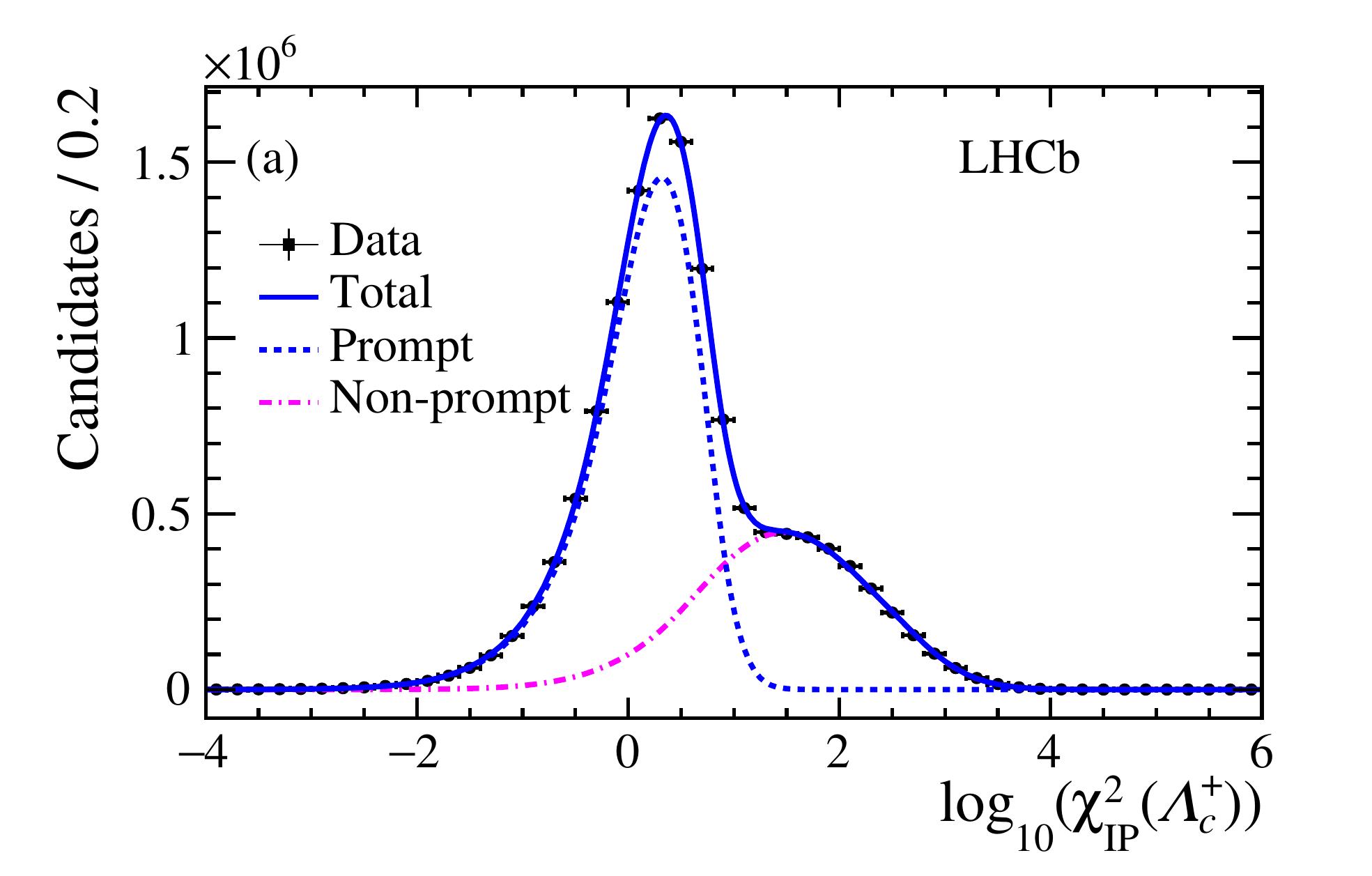}
\end{minipage}
\begin{minipage}[t]{0.49\textwidth}
\centering
\includegraphics[width=1.0\textwidth]{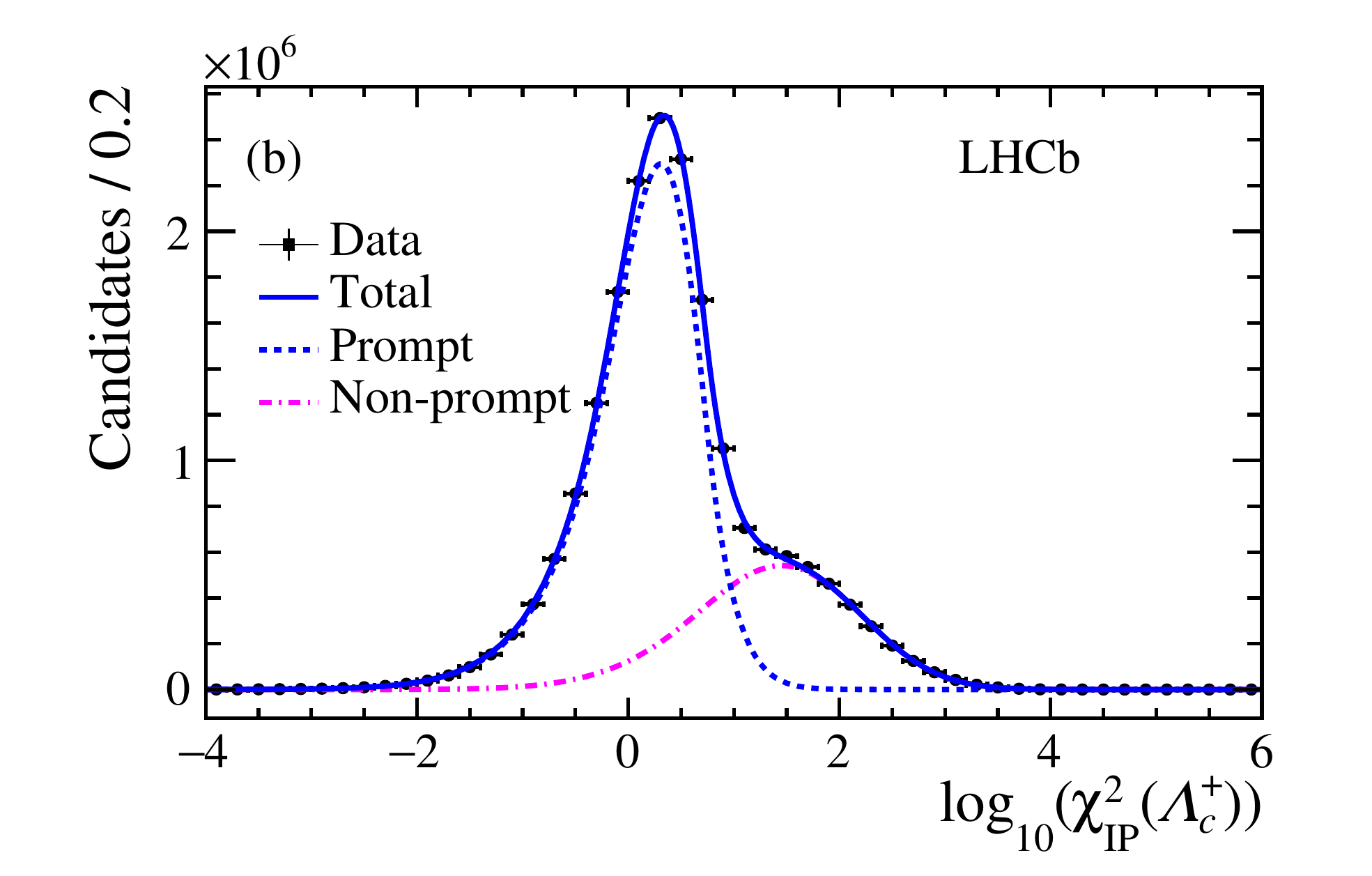}
\end{minipage}
\caption{Distributions of $\log_{10}(\chisqip(\Lc))$ for background-subtracted candidates (a)
triggered by TOS and (b) triggered by exTIS,
with fit results shown.}
\label{fig:LcIP}
\end{figure}

The determination of the prompt $\Lcp$ baryon yields, which are contaminated by $\Lcp$ candidates produced in $b$-hadron decays, is done in two steps~\cite{LHCb-PAPER-2012-041}.
First, a binned extended maximum-likelihood fit to the $m(pK^-\pip)$ invariant-mass distribution in the interval 
2220--2360\mevcc
is performed to determine the total number of $\Lcp$ candidates. 
Then a binned extended maximum-likelihood fit to the
background-subtracted   
${\rm log}_{10}(\chisqip(\Lc))$ distribution
is performed to separate the prompt $\Lcp$ component from that originated in $b$-hadron decays. 
The mass distribution of $\Lcp$ candidates is described by a sum of
a Gaussian function and  a modified Gaussian function with power-law
tails on both sides with a common peak position.
The background mass distribution is described
by a first-order Chebyshev polynomial.
The ${\rm log}_{10}(\chisqip(\Lc))$ distribution, after subtracting the combinatorial background 
using the $sPlot$ technique~\cite{Pivk:2004ty},
is described by  
two Bukin functions~\cite{Bukin:2007}.
All the parameters except the peak position and resolution of the
functions are derived from a fit to simulated signal.
Figures~\ref{fig:LcMass} and~\ref{fig:LcIP} 
show the $p K^- \pip$ invariant-mass distribution 
and
${\rm log}_{10}(\chisqip(\Lc))$ distributions 
in data together with the 
fit results for the two
trigger categories.
The signal yields for both the signal and the normalisation modes 
are presented in Table~\ref{tab:sigresults}.

%%%%%%%%%%%%%%%%%%%%%%%%%%%%%%%%%%%%%%%%%%%%%%%%%%%%%%%%%%e
%%%%%%%%%%%%%%%%%%%%%%%%%%%%%%%%%%%%%%%%%%%%%%%%%%%%%%%%%%
%%%%%%%%%%%%%%%%%%%%%%%%%%%%%%%%%%%%%%%%%%%%%%%%%%%%%%%%%%e
%%%%%%%%%%%%%%%%%%%%%%%%%%%%%%%%%%%%%%%%%%%%%%%%%%%%%%%%%%
%%%%%%%%%%%%%%%%%%%%%%%%%%%%%%%%%%%%%%%%%%%%%%%%%%%%%%%%%%e
%%%%%%%%%%%%%%%%%%%%%%%%%%%%%%%%%%%%%%%%%%%%%%%%%%%%%%%%%%
\begin{table}
\caption{Yields of the signal and normalisation modes.}
\centering
\begin{tabular}{ >{\small}l|>{\small}c|>{\small}c }
\toprule
Category & $N_{\text{sig}}$ & $N_{\text{norm}} [10^{3}] $ \\
\midrule
TOS &  $116 \pm 23$&\phantom{0}$8764 \pm 6$  \\
exTIS &$210 \pm 29$& $13889 \pm 8$ \\ 
\bottomrule
\end{tabular}\label{tab:sigresults}
\end{table}

\section{Efficiencies}
\label{sec:eff}
For each trigger category and for both the signal and the normalisation channels,
the total efficiencies
are computed as products
of the detector geometrical 
acceptance and of the efficiencies related to particle reconstruction,
event selection, particle identification and trigger. 
All the efficiencies are calculated
using simulation that is corrected using data. 
For both the signal and the normalisation modes, 
the kinematic distributions in simulation samples, including the transverse momentum and rapidity of the $\Xiccpp$ and $\Lcp$ baryons and the event multiplicity, are weighted to match those in the corresponding data. 
The efficiencies are calculated under three lifetime ($\tau_{\Xiccpp}$) hypotheses: 
the central value of the measured lifetime, and the lifetime increased or decreased by its measured uncertainty~\cite{LHCb-PAPER-2018-019}.
The dependence of the efficiency on the \Xiccpp baryon lifetime is almost linear, with the efficiency ratio varying by 25\% from the lower lifetime to the higher one.
The resonant structures of the $\Lcp \to \proton \Km \pip$ decay are also weighted based on
the background-subtracted data,
as the simulation samples do not model well the structure seen in the data.
The tracking efficiency is corrected with control data samples, as described in Ref.~\cite{LHCb-DP-2013-002}.
The particle-identification efficiency is corrected in bins of particle momentum, pseudorapidity and event multiplicity, using the results of a tag-and-probe method applied to calibration samples~\cite{LHCb-DP-2018-001}.
The efficiency ratios of the normalisation mode to the signal mode
are presented in Table~\ref{tab:effresults}.
\begin{table}
\caption{Ratios of the normalisation and signal efficiencies.}
\centering
\begin{tabular}{>{\small}l|>{\small}c|>{\small}c|>{\small}c }
\toprule
 &\multicolumn{3}{c}{$\varepsilon_{\text{norm}}$/$\varepsilon_{\text{sig}}$} \\
\cline{2-4}
Category& $\tau_{\Xiccpp}$ = 0.230 \ps &$\tau_{\Xiccpp}$ = 0.256 \ps &$\tau_{\Xiccpp}$ = 0.284 \ps \\
\midrule
TOS &$22.00 \pm 1.09$ & $19.50 \pm 1.71$ &$17.50 \pm 1.50$  \\
exTIS &$16.64 \pm 1.30$ & $14.56 \pm 1.06$ &$12.95 \pm 0.80$ \\ 
\bottomrule
\end{tabular}\label{tab:effresults}
\end{table}

\section{Systematic uncertainties}
\label{sec:sys}
The sources of systematic uncertainties affecting the measurement of the production ratio  
include
the choice of the fit model and 
the evaluation of the total efficiency.
The uncertainties are summarised in Table~\ref{tab:BRsyserrTOS}.

For both the signal and normalisation modes, the uncertainties due to the choice of the particular fit model are estimated by 
using alternative functions where the signal is described by 
a sum of two Gaussian functions with a common peak position
and the background is
described by a second-order polynomial function.
The difference in the ratio of signal yields between the two fits is assigned as systematic uncertainty.
Additional effects
coming from the $\log_{10}(\chisqip(\Lc))$ fit are tested with alternative functions 
where the parameters 
used to describe the nonprompt signal
are determined from a $\Lb$ baryon data sample. 
The effect from the background subtraction is studied using the shape
determined with the candidates in the $\Lcp$ baryon mass sidebands. 

The limited size of the simulation
samples leads to systematic uncertainties on the efficiencies.
The systematic uncertainty due to the trigger selection efficiency is
estimated with a tag-and-probe method 
exploiting a sample of events that are also triggered by particles unrelated to the signal
candidate~\cite{LHCb-DP-2012-004}.
Due to the small sample size of the signal channel in data, two different control samples are used. 
The first sample comprises 
$\Lb \to \Lc \pi^ -\pi^+ \pi^-$ decays, which are topologically similar to the $\Xiccpp \to \Lcp\Km\pip\pip$ decay. 
The second 
sample comprises $\Bc\to\jpsi\pi^+$ 
decays. This decay does not have the same topology but shares another feature with the signal: there should be at least two other heavy-flavour particles (\bquark- or \cquark-hadrons) produced in the same event that can be responsible for the trigger decision. 
The hardware trigger efficiencies
of the 
$\Lb$, $\Bc$ decay channels and prompt $\Lc$ channel,
are measured using the tag-and-probe method.
Similar selections to those applied
to the signal channel are applied to
both the data and simulation for the control samples. 
The efficiency ratio of the $\Lb$, $\Bc$ decays to the $\Lcp$ decays is estimated and 
the difference of the ratio in data and in simulation is assigned as a
systematic uncertainty. 
The transverse-energy threshold in the calorimeter hardware trigger 
varied during data taking, and this variation is
not fully described by the simulation.
The threshold used in the simulated samples is higher than that applied to some data. 
To investigate the
influence of this difference, the same hardware trigger requirement used in the simulation
is applied to the data.
The measurement is repeated and the change in the measured production ratio is taken as a systematic uncertainty.

The systematic uncertainty related to the tracking efficiency
includes three effects.
First, the tracking efficiency depends on the detector occupancy, which is not well described by simulation.  
The distribution of the number of SPD hits in simulated samples
is weighted to match that in data and an uncertainty of $0.8\%$ per track is
assigned to account for  remaining difference in multiplicity  between data and simulation~\cite{LHCb-DP-2013-002}. 
Secondly, the uncertainty due to the finite size of the control samples is
propagated to the final systematic uncertainty using a large number of pseudoexperiments.
Finally, an uncertainty is assigned to 
the track reconstruction efficiency due to uncertainties on the material budget of the
detector and on the modelling of hadronic interaction with the detector material.

The systematic uncertainty related to the particle-identification efficiency
includes three effects.
The effect from the limited size of calibration samples is evaluated
with a large number of pseudoexperiments. Effects of 
binning in momentum, pseudorapidity and event multiplicity
is evaluated by increasing or decreasing the bin sizes by a factor of two.
In this estimation, the effects of the correlations between tracks on the particle identification performance are taken into account using simulated samples.

The uncertainties on the weights used for the correction of the kinematic distributions of the simulation samples are propagated as a systematic uncertainty on the production ratio.

%%%%%%%%%%%%%%%%%%%%%%%%%%%%%%%%
\begin{table}
    \caption{Relative systematic uncertainties on the production ratio measurement for the two trigger categories.}
\centering
\begin{tabular}{l|c|c}
\toprule
Source &TOS  $[\%]$ &exTIS $[\%]$ \\
\midrule
Simulation sample size  &\phantom{0}8.8 &\phantom{0}7.3\\
Fit model&\phantom{0}5.4 & \phantom{0}5.3\\ 
Hardware trigger & \phantom{0}9.0  &\phantom{0}6.3\\
Tracking  &\phantom{0}3.4 &\phantom{0}3.4 \\
Particle identification &\phantom{0}5.5 &\phantom{0}5.4  \\
Kinematic correction &\phantom{0}7.3 &\phantom{0}6.0 \\
\hline
Sum in quadrature &16.8 &14.1\\
\bottomrule
\end{tabular}\label{tab:BRsyserrTOS}
\end{table}

\section{Results}
The production-rate ratio is calculated for the
TOS and the exTIS categories of events for three different $\Xiccpp$ lifetime
scenarios using Eq.~(\ref{eq:strategy:defineMainResult}). 
The separate ratios in the TOS and exTIS categories are presented in Table~\ref{tab:results} and
are found to be consistent. 
The combination of the trigger categories, using the Best Linear Unbiased Estimate method~\cite{Nisius:2014wua,*blue} is also reported.
In the combination, the systematic uncertainties coming from the simulation sample size and hardware trigger are assumed to be uncorrelated, while the other systematic uncertainties are considered to be 100$\%$ correlated.
\begin{table}
    \caption{Production rate ratio results 
for three different $\Xiccpp$ lifetime hypotheses. The first uncertainty is statistical and the second is systematic.}
\centering
\begin{tabular}{ >{\small}l|>{\small}c|>{\small}c|>{\small}c }
\toprule
 & \multicolumn{3}{c}{$R\; [10^{-4}]$} \\
\cline{2-4}
Category & $\tau_{\Xiccpp}$ = 0.230 \ps &$\tau_{\Xiccpp}$ = 0.256 \ps &$\tau_{\Xiccpp}$ = 0.284 \ps \\
\midrule
TOS & $\XiccTOSl$ & $\XiccTOS$ &$\XiccTOSh$ \\
exTIS &$\XiccTISl$ & $\XiccTIS$ &$\XiccTISh$\\ 
\hline
Combined&  $\XiccComl$ & $\XiccCom$ & $\XiccComh$\\
\bottomrule
\end{tabular}\label{tab:results}
\end{table}
%%%%%%%%%%%%%%%%%%%%%%%%%%%%%%%%

\section{Conclusion}\label{sec:conclusion}
A first measurement of the $\Xiccpp$ production cross-section relative to that of \Lcp  baryons is presented.
The ratio of $\Xiccpp$ production cross-section times
  the branching fraction of the  $\Xiccpp \to \Lc K^-
  \pi^+ \pi^+$ decay relative to the prompt $\Lc$ production
  cross-section in the kinematic region $4<\pt<15\gevc$ and $2.0<y<4.5$
  is measured to be 
  $(2.22\pm 0.27 \pm 0.29)\times 10^{-4}$, assuming the central value of the $\Xiccpp$ lifetime measured in Ref.~\cite{LHCb-PAPER-2018-019},
  where the first uncertainty
  is statistical and the second systematic.
This is the first measurement of the production of the doubly charmed baryons in $pp$ collisions and will deepen our understanding on their production mechanism.

\section*{Acknowledgements}
%
% These Acknowledgements valid from 3-May-2019
%
\noindent We thank Chao-Hsi Chang, Cai-Dian L\"u, Xing-Gang Wu, and Fu-Sheng Yu for the discussions on the production and decays of double-heavy-flavour baryons.
We express our gratitude to our colleagues in the CERN
accelerator departments for the excellent performance of the LHC. We
thank the technical and administrative staff at the LHCb
institutes.
We acknowledge support from CERN and from the national agencies:
CAPES, CNPq, FAPERJ and FINEP (Brazil); 
MOST and NSFC (China); 
CNRS/IN2P3 (France); 
BMBF, DFG and MPG (Germany); 
INFN (Italy); 
NWO (Netherlands); 
MNiSW and NCN (Poland); 
MEN/IFA (Romania); 
MSHE (Russia); 
MinECo (Spain); 
SNSF and SER (Switzerland); 
NASU (Ukraine); 
STFC (United Kingdom); 
DOE NP and NSF (USA).
We acknowledge the computing resources that are provided by CERN, IN2P3
(France), KIT and DESY (Germany), INFN (Italy), SURF (Netherlands),
PIC (Spain), GridPP (United Kingdom), RRCKI and Yandex
LLC (Russia), CSCS (Switzerland), IFIN-HH (Romania), CBPF (Brazil),
PL-GRID (Poland) and OSC (USA).
We are indebted to the communities behind the multiple open-source
software packages on which we depend.
Individual groups or members have received support from
AvH Foundation (Germany);
EPLANET, Marie Sk\l{}odowska-Curie Actions and ERC (European Union);
ANR, Labex P2IO and OCEVU, and R\'{e}gion Auvergne-Rh\^{o}ne-Alpes (France);
Key Research Program of Frontier Sciences of CAS, CAS PIFI, and the Thousand Talents Program (China);
RFBR, RSF and Yandex LLC (Russia);
GVA, XuntaGal and GENCAT (Spain);
the Royal Society
and the Leverhulme Trust (United Kingdom).

\addcontentsline{toc}{section}{References}
\setboolean{inbibliography}{true}
\bibliographystyle{LHCb}
\bibliography{main.bbl} 

\newpage
 
\newpage
% LHCb collaboration author list
% Data extracted on December 14th, 2019 at 9:49pm for reference date 09-Dec-2019
\centerline
{\large\bf LHCb collaboration}
\begin
{flushleft}
\small
R.~Aaij$^{31}$,
C.~Abell{\'a}n~Beteta$^{49}$,
T.~Ackernley$^{59}$,
B.~Adeva$^{45}$,
M.~Adinolfi$^{53}$,
H.~Afsharnia$^{9}$,
C.A.~Aidala$^{80}$,
S.~Aiola$^{25}$,
Z.~Ajaltouni$^{9}$,
S.~Akar$^{66}$
% ...

{\footnotesize \it

% ...
$ ^{9}$Universit{\'e} Clermont Auvergne, CNRS/IN2P3, LPC, Clermont-Ferrand, France\\
% ...
$ ^{25}$INFN Sezione di Milano, Milano, Italy\\
% ...
$ ^{31}$Nikhef National Institute for Subatomic Physics, Amsterdam, Netherlands\\
% ...
$ ^{45}$Instituto Galego de F{\'\i}sica de Altas Enerx{\'\i}as (IGFAE), Universidade de Santiago de Compostela, Santiago de Compostela, Spain\\
% ...
$ ^{49}$Physik-Institut, Universit{\"a}t Z{\"u}rich, Z{\"u}rich, Switzerland\\
% ...
$ ^{53}$H.H. Wills Physics Laboratory, University of Bristol, Bristol, United Kingdom\\
% ...
$ ^{59}$Oliver Lodge Laboratory, University of Liverpool, Liverpool, United Kingdom\\
% ...
$ ^{66}$Los Alamos National Laboratory (LANL), Los Alamos, United States\\
$ ^{67}$Syracuse University, Syracuse, NY, United States\\
% ...
$ ^{80}$University of Michigan, Ann Arbor, United States, associated to $^{67}$\\
\bigskip
% ...
\medskip
%$ ^{\dagger}$Deceased
}
\end{flushleft}

\end{document}